%% file: main.tex
\documentclass[conference,9pt]{IEEEtran}
\IEEEoverridecommandlockouts

\usepackage[utf8]{inputenc} 
\usepackage{pifont} 
\usepackage{xcolor}
\usepackage{times}
\usepackage{graphicx}
\usepackage{booktabs}
\usepackage{algorithm}
\usepackage{algpseudocode}
\usepackage[compress]{cite}
\usepackage{amsmath,amsfonts,amssymb}
\usepackage{mathtools}
\usepackage{multirow}
\usepackage{threeparttable}
\usepackage{subcaption}
\usepackage{makecell}
\usepackage{footnote}
\usepackage{kotex}
\usepackage{url}
\usepackage{flushend}
\usepackage{adjustbox}
\usepackage[normalem]{ulem}

\input{cmds}

\graphicspath{{figs/}}
\newcommand{\jlnote}[1]{{\color{orange}\em [Jongeun's note: #1]}}
\newcommand{\jwnote}[1]{{\color{blue}\em [JW note: #1]}}
\newcommand{\sysname}{EP-HDC\xspace}

\begin{document}

\title{
EP-HDC: Hyperdimensional Computing with Encrypted Parameters for High-Throughput Privacy-Preserving Inference

\thanks{$^\ddagger$J.~Lee is the corresponding author.}
\thanks{This work was partly supported by the NRF Grant through National R\&D Program (RS-2024-00360300, 50\%), by the IITP Grant through Artificial Intelligence Graduate School Program (UNIST, RS-2020-II201336, 10\%), and by the IITP-ITRC grant (IITP-2025-RS-2025-II211817, 40\%), all funded by the Korea government. The EDA tool was supported by the IC Design Education Center (IDEC), Korea.}
}

\author{%
  \IEEEauthorblockN{%
    Jaewoo Park\IEEEauthorrefmark{1},
    Chenghao Quan\IEEEauthorrefmark{2} and
    Jongeun Lee\IEEEauthorrefmark{2}\IEEEauthorrefmark{3}
  }%
 \IEEEauthorblockA{\IEEEauthorrefmark{1}Department of Computer Science and Engineering,
 \IEEEauthorrefmark{2}Department of Electrical Engineering}%
\IEEEauthorblockA{Ulsan National Institute of Science and Technology (UNIST), Ulsan, Korea}%
\IEEEauthorblockA{\texttt{\{hecate64,quanch,jlee\}@unist.ac.kr}}%
\vspace{-5mm}
}

\maketitle


\input{body}

\bibliographystyle{IEEEtran}
\bibliography{refs}

\end{document}

%% file: cmds.tex
\usepackage{xspace}

\newcommand{\x}{$\times$\xspace}
\newcommand{\X}{{\footnotesize $\times$}\xspace}

\newcommand{\ra}{$\rightarrow$\xspace}

\newcommand{\ceil}[1]{\ensuremath{\left\lceil#1\right\rceil}}
\newcommand{\floor}[1]{\ensuremath{\left\lfloor#1\right\rfloor}}

\usepackage{array}
\newcolumntype{P}[1]{>{\centering\arraybackslash}p{#1}}

\newcommand{\figref}[1]{Fig.~\ref{#1}}
\renewcommand{\algref}[1]{Algorithm~\ref{#1}}
\newcommand{\secref}[1]{Section~\ref{#1}}

\newcommand{\tabref}[1]{Table~\ref{#1}}

\newcommand{\ignore}[1]{}
\newcommand{\etal}{\textit{et al.}\xspace}

\newcommand{\red}[1]{{\color{red} #1}}

\newcommand{\myincludegraphics}[3][]{\centering
\includegraphics[#1]{#2}\caption{#3}\label{#2}\bottomoffigure}

\newcommand{\bottomoffigure}{}

\usepackage{amsthm}
\makeatletter\newcommand{\ppheader}[2][1mm]{{\ifx\x#1\x{}\else{\vspace{#1}}\fi}\noindent\textbf{#2}\@addpunct{.}~}\makeatother

\newcounter{todocounter}
\makeatletter\@ifpackageloaded{todonotes}{\newcommand{\todonum}[2][]{\stepcounter{todocounter}\todo[inline]{\thetodocounter: #2\ifx\x#1\x{}\else{ [#1]}\fi}}}{\newcommand{\todonum}[2][]{\par\stepcounter{todocounter}\noindent\fcolorbox{black}{green!20}{\begin{minipage}{.98\linewidth}\thetodocounter: #2\ifx\x#1\x{}\else{ [#1]}\fi\end{minipage}}}}\makeatother

\makeatletter\@ifclassloaded{IEEEtran}
{
\newenvironment{keyword}{\begin{IEEEkeywords}}{\end{IEEEkeywords}}
\ifCLASSOPTIONconference\fi}{
\def\bstctlcite{\@ifnextchar[{\@bstctlcite}{\@bstctlcite[@auxout]}}
\def\@bstctlcite[#1]#2{\@bsphack\@for\@citeb:=#2\do{%
\edef\@citeb{\expandafter\@firstofone\@citeb}%
\if@filesw\immediate\write\csname #1\endcsname{\string\citation{\@citeb}}\fi}%
\@esphack}}\makeatother

%% file: body.tex



\begin{abstract}
  While homomorphic encryption (HE) provides strong privacy protection, its high computational cost has restricted its application to simple tasks. Recently, hyperdimensional computing (HDC) applied to HE has shown promising performance for privacy-preserving machine learning (PPML). However, when applied to more realistic scenarios such as batch inference, the HDC-based HE has still very high compute time as well as high encryption and data transmission overheads.
  To address this problem, we propose \emph{HDC with encrypted parameters (\sysname)}, which is a novel PPML approach featuring \emph{client-side HE}, i.e., inference is performed on a client using a homomorphically encrypted model. Our \sysname can effectively mitigate the encryption and data transmission overhead, as well as providing high scalability with many clients while providing strong protection for user data and model parameters. In addition to application examples for our client-side PPML, we also present design space exploration involving quantization, architecture, and HE-related parameters.
  Our experimental results using the BFV scheme and the Face/Emotion datasets demonstrate that our method can improve throughput and latency of batch inference by orders of magnitude over previous PPML methods (36.52$\sim$1068\X and 6.45$\sim$733\X, respectively) with $<$1\% accuracy degradation.

\end{abstract}

\maketitle

\section{Introduction}
As artificial intelligence gains popularity, there is an increasing demand to ensure privacy of user data. Homomorphic Encryption (HE) offers a robust solution by enabling arithmetic operations on ciphertext without ever revealing the encryption key or plaintext. However, HE-based privacy preserving machine learning (PPML) has been restricted to very small networks and simple tasks due to the high computational cost and the limitation on the supported operations in HE \cite{cryptonets:icml16,lola}.

Recently, a promising alternative based on hyperdimensional computing (HDC) has emerged \cite{hehdc}. 
HDC performs inference in two steps, and the latter (i.e., similarity search) involves only a few simple arithmetic operations that can be very efficiently mapped to HE.  Consequently, HE-evaluated HDC, or \emph{HE-HDC} for short, has been shown to deliver orders of magnitude faster performance than Deep Neural Network (DNN)-based PPML methods \cite{hehdc}.
However, in batch inference scenarios (i.e., when multiple input queries are available for processing), the previous HE-HDC approach suffers from relatively high encryption and data transmission overhead as well as high HE compute time, limiting its efficiency significantly. 

To overcome this problem, we propose \emph{hyperdimensional computing with encrypted parameters (\sysname)}, which is a novel PPML approach featuring \emph{client-side HE}, meaning that it runs inference on a client using a homomorphically encrypted model.
The key insight is that while PPML requires either a model or input data to be encrypted, a model may be much smaller than input data, and therefore much cheaper to protect, which is indeed the case with HDC-based PPML.
Our \sysname dramatically reduces the encryption and data transmission overhead, is significantly more efficient on batch inference with no downside in inference accuracy compared to HE-HDC, and provides high scalability with many clients, while still providing strong protection for user data and model parameters. We present application examples for \sysname as well as parameter optimization through extensive design space exploration involving quantization, architecture, and HE parameters.

Our experimental results using the MNIST, Face \cite{face}, and Emotion \cite{emotion} datasets demonstrate that our \sysname can achieve 612\X higher throughput compared with the previous best HDC-based PPML (HE-HDC) \cite{hehdc} without compromising inference accuracy.  Compared with DNN-based PPML methods, our \sysname achieves 36.52$\sim$1068\X higher throughput and 6.45$\sim$733\X lower latency with $<$1\% accuracy degradation.

Our proposed method is based on HDC.  Though presently HDC is restricted to simpler tasks, current DNN-based PPML also faces scalability issues due to excessive runtime and the polynomial approximation of nonlinear operations (e.g., ReLU approximated to square \cite{lola,cryptonets:icml16}).  Thus, HDC-based PPML can offer a viable alternative solution for applications where both privacy and performance are crucial.



We make the following contributions in this paper:
\begin{itemize}
\item We propose a client-side PPML method, \sysname, with a security analysis and application examples.
\item We perform extensive design space exploration of \sysname involving quantization, architecture, and HE parameters. 
\item We evaluate the performance of \sysname on commodity hardware, demonstrating its efficiency over previous PPML methods.
\end{itemize}
\vspace{-0.8mm}

\begin{figure*}
  \myincludegraphics[width=.8\linewidth]{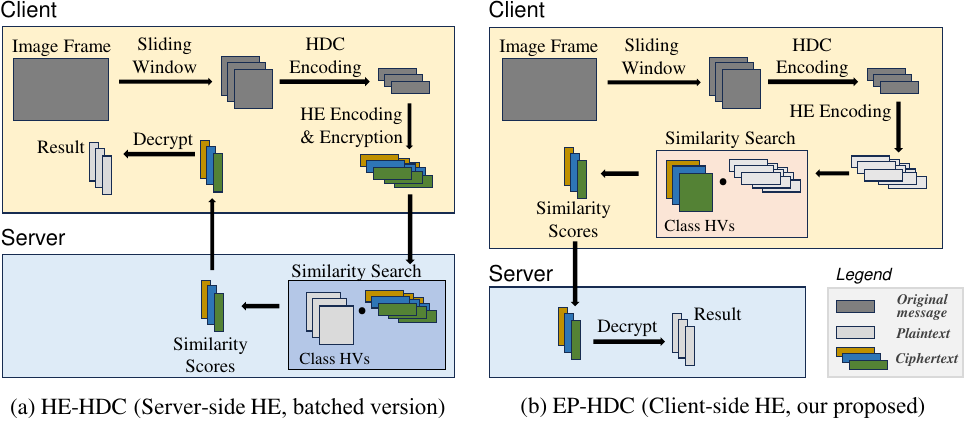}{HE-HDC vs.\ the proposed \sysname for batch HDC inference. HV stands for hypervector.}
  \vspace{-3mm}
\end{figure*}

\section{Preliminary}

\subsection{Hyperdimensional Computing}

HDC is a brain-inspired machine learning paradigm based on high-dimensional vector representation called \emph{hypervector} \cite{kanerva2009hyperdimensional}.
The HDC approach has many advantages over DNNs including more efficient training (not based on back-propagation), higher robustness against noise, and superior generalization performance \cite{rahimi2016hyperdimensional, 7838428, thomas2021theoretical, neuralhd}.



\ppheader{HDC Encoding}
The goal of HDC encoding is to discover a well-defined hypervector representation for input data. Though there are multiple existing encoding schemes \cite{neuralhd, voicehd, adapthd, onlinehd}, they all satisfy the common-sense principle: different data points in the original space should have orthogonal hypervectors in the HDC space. For example, random projection encoding~\cite{onlinehd} transforms a flattened image vector into hypervectors by multiplying a random base hypervector. 

\ppheader{Single-pass Training}
Once input hypervectors are generated, the hypervectors
belonging to the same class are averaged to obtain a class hypervector. 

\ppheader{Inference}
The classification of a query data is performed by first encoding the query data into a query hypervector $\vec{\mathcal{H}}_q$, and then computing the similarity of $\vec{\mathcal{H}}_q$ with each class hypervector. 
The class with the highest similarity score is returned as the classification result. For the similarity measure, cosine similarity is widely used.

\ignore{\ppheader{Iterative Training (Retraining)}
Due to the poor classification accuracy of single-pass training, using an iterative training method is suggested, which discards mispredicted queries from mispredicted classes and adds them to correct classes \cite{onlinehd}. 
}

\subsection{Homomorphic Encryption}

Homomorphic encryption allows arithmetic operations such as multiplication and addition to be evaluated on encrypted data. The majority of HE schemes are based on the ring learning with errors (LWE) problem, where the underlying operations are performed over polynomials of degree $N$. In HE, a message is first \emph{encoded} into a plaintext $(c_0,c_1)$, which is a pair of polynomials of degree $N$, and then the plaintext is \emph{encrypted} into a ciphertext using a given secret key. Homomorphic operations can be performed between two ciphertexts or between a ciphertext and a plaintext, resulting in a ciphertext encrypted with the same secret key.

HE schemes such as CKKS \cite{ckks}, BGV \cite{bgv}, and BFV \cite{bfv} can encode a vector of $N$ integers into a single plaintext ($N/2$ fixed-points for CKKS), which is referred to as \emph{packing}. When packing is used, homomorphic multiplication and addition are evaluated as element-wise operations between two message vectors. However, the individual elements of a ciphertext message vector cannot be accessed unless decrypted. The only way to change the position of elements inside the message vector is by homomorphic rotation. 
Homomorphic rotation has a much longer latency than addition and ciphertext-plaintext multiplication, and therefore should be avoided as much as possible to achieve high performance.
\ignore{
\begin{table}[t]
  \centering
  \caption{Comparison between different HE schemes}
  \label{t:scheme_comp}
  \begin{tabular}{c|ccc}
      \toprule
                 & CKKS                & BGV     & BFV     \\
                     \midrule
Arithmetic       & fixed-point complex & integer & integer \\
\# of SIMD slots & $N/2$                 & $N$       & $N$      \\
    \bottomrule
\end{tabular}
\end{table}}


\subsection{Message Representation in HE-based PPML}
In HE-based PPML, input vectors can be packed in various ways into plaintext.
CryptoNets \cite{cryptonets:icml16} uses a packing where each element of a vector results in a different plaintext message, but multiple vectors (from multiple input images) share the same set of messages, which can be called \emph{SIMD representation}.  With the SIMD representation, mapping computation to HE is straightforward, but it requires a large number of inputs to fill the SIMD slots, which can be quite many (e.g., 4K).  To reduce latency, LoLa \cite{lola} advocates different message representations such as \emph{dense representation}, where one input vector is represented by one plaintext message. However, this requires more complicated mapping.  For instance, the dot-product of two vectors in dense representation needs one multiplication followed by $\log_2 N$ HE rotations (which are expensive), where $N$ is the polynomial degree. 

\subsection{Threat Model}
This work assumes the common threat model suggested in previous PPML works \cite{cryptonets:icml16,lola,hehdc}. This model involves two parties, the client and the server, with the server providing ML-based object detection as a service. The client does not want to expose its original image to the server, while the server prefers not to reveal the ML model (i.e., the class hypervectors) to the client. Similar to two-party computation \cite{reagen2021cheetah,gazelle}, we assume that the client is honest in performing its computation but curious about the ML model.

\begin{table}
\centering
\caption{Comparison of PPML Methods}
\label{t:comparison}
\begin{threeparttable}
\begin{tabular}{cl|lll}
  \toprule
  Alg. & PPML Method & Msg Rep. & Batch Size & HE Compute \\
  \midrule
  \multirow{2}{*}{DNN} & CryptoNets \cite{cryptonets:icml16} & SIMD & Large (4K) & Server \\
  &  LoLa \cite{lola} & Multiple\tnote{*} & 1 & Server \\
  \midrule
 & HE-HDC\cite{hehdc} & Dense & 1 & Server \\
HDC  & HE-HDC-SR & SIMD & Large & Server \\
   & EP-HDC & SIMD & Large & Client\\
  \bottomrule
\end{tabular}
\begin{tablenotes}
\item[*] Using multiple representations incl.\ dense, interleave, convolution, etc.
\end{tablenotes}
\end{threeparttable}
\end{table}

\ignore{
\begin{table}
\centering
\caption{Comparison of PPML Methods on MNIST Inference}
\label{t:comparison}
\begin{threeparttable}
\resizebox{\linewidth}{!}{%
\begin{tabular}{cl|lr@{~/~}rr}
  \toprule
  Alg. & PPML Method & Msg Rep. & Latency & Batch & Accuracy \\
  \midrule
  \multirow{2}{*}{DNN} & CryptoNets \cite{cryptonets:icml16} & SIMD & $\sim$250 s & 4096 & 98.95\% \\
  &  LoLa \cite{lola} & Multiple\tnote{1} & 2.2 s\tnote{2} & 1 & 98.95\% \\
  \midrule
 \multirow{3}{*}{\makecell{HDC \\ (D=4K)}} & HE-HDC\cite{hehdc} & Dense & 84 ms & 1 & 97.10\% \\
   & HE-HDC-SR & SIMD & 4599 ms\tnote{3} & 204 & 97.10\% \\
   & EP-HDC & SIMD & 408 ms & 204 & 98.39\%  \\
  \bottomrule
\end{tabular}}
\begin{tablenotes}
\item[1] Using multiple representations incl.\ dense, interleave, convolution, etc.
\item[2] Multithreading enabled
\item[3] Improvement mainly due to eliminating HE rotations
\end{tablenotes}
\end{threeparttable}
\end{table}
}

\begin{table}
  \caption{List of Symbols}
  \label{t:sym}
  \centering
  \begin{tabular}{c|l}
    \toprule
    Symbol & \multicolumn{1}{c}{Description} \\
    \midrule
    $N$ & Polynomial degree for plaintext/ciphertext (a power of 2) \\
    $\log_2 t$ & Bit-length of a message (each element's in case of a vector) \\
    $D$ & Hypervector dimension (typ. 4K$\sim$10K) \\
    $k$ & Number of output classes (or categories) \\
    \bottomrule
  \end{tabular}
  \vspace{-5mm}
\end{table}

\ignore{\begin{figure}
  \myincludegraphics[width=\linewidth]{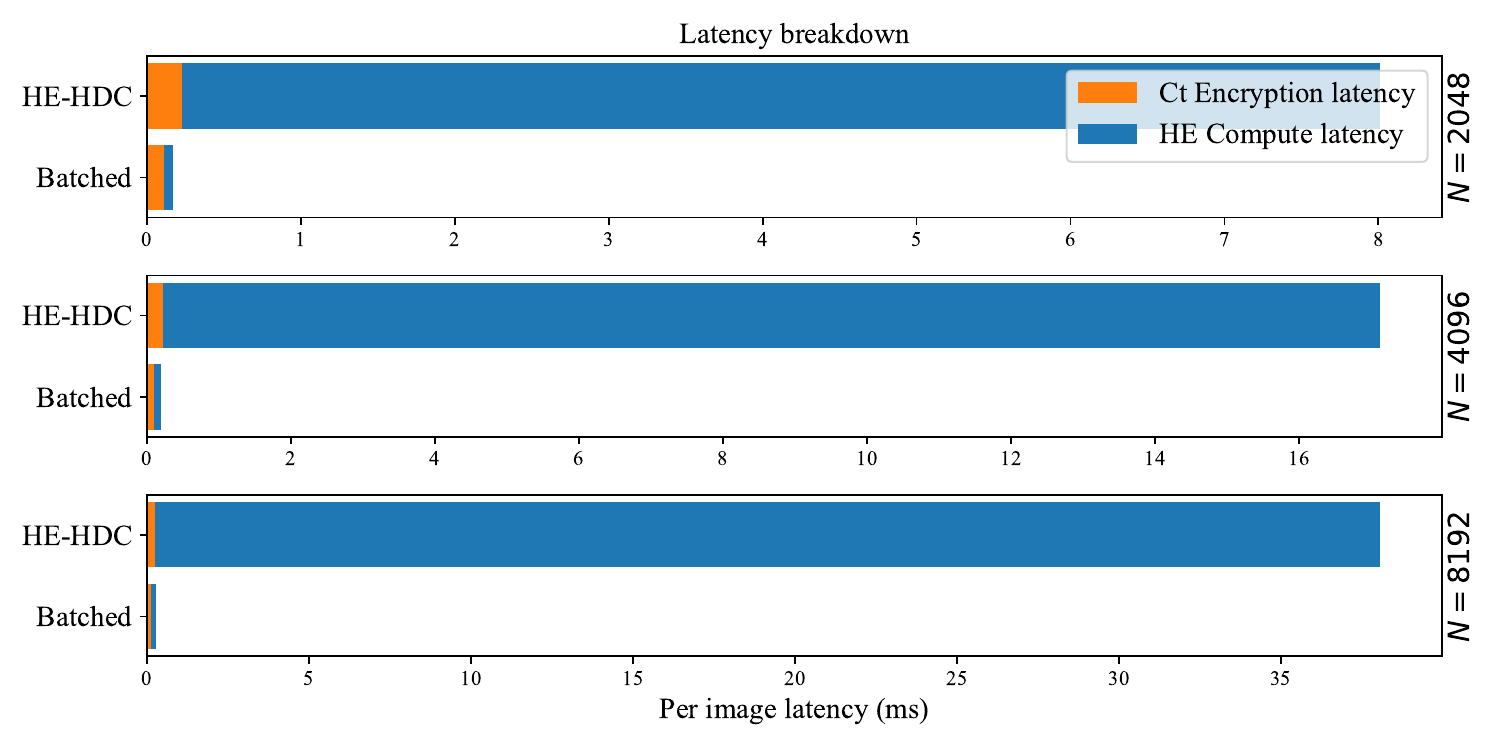}{Runtime breakdown of server-side HE (see also \figref{overview}).}
  \jwnote{Preliminary plot, I will receive feedback and update the figure. }
  \jwnote{The per-image latency is plotted (total latency / batch size). HE-HDC has batch size =1 and the 'Batched' case has a batch size of (N/\#of classification labels). }
  // per-image lat?  // not very interesting figure?  // why only two segs?
\end{figure}
}


\section{Analysis of Batched HE-HDC}
In batch inference, a number of input queries are sent together to the server so that the server can process them simultaneously, potentially gaining processing efficiency.
For instance, in computer vision, one way to do object detection is to first generate a number of image patches, also called proposals,  and then to perform image classification on those patches of image.  Face detection and recognition also can be performed similarly.
Obviously such applications have much higher computational demand than simple image classification, necessitating new approaches to improve efficiency significantly.

The HDC-based PPML approach (e.g., HE-HDC \cite{hehdc}) can provide much higher performance than DNN-based PPML methods.  The idea of HE-HDC is to (i) replace DNN inference with HDC inference, and (ii) perform the similarity search of HDC on the server using encrypted query hypervectors (HVs), while HDC encoding (i.e., \emph{encoding} image patches into query HVs) is performed by the client in the plaintext domain, as illustrated in \figref{overview}(a).  Note that the client must also perform the \emph{encryption} of query HVs and the decryption of similarity scores.

While HE-HDC can be orders of magnitude faster than DNN-based PPML methods, HE-HDC has certain weaknesses in batch inference. HE-HDC is essentially a \emph{sequential} approach in the sense that it processes one query HV at a time regardless of the number of input queries available, due to the use of the \emph{dense representation} \cite{lola}. 
%
%
Alternatively, we can take a parallel approach; instead of encrypting one query HV to one ciphertext\footnote{Or multiple ciphertexts if $D > N$ where $D$ is the size (or dimension) of an HV, and $N$ is the number of SIMD slots of an HE scheme.}, we can encrypt $N$ HVs into $D$ ciphertexts  by utilizing the SIMD representation (note, each HV is a $D$-dim vector, and each ciphertext has $N$ slots). This can eliminate the expensive HE rotations in the dot-product computation, resulting in significantly improved throughput (8$\sim$11\X depending on $N$) compared to the serial version.
\tabref{t:comparison} compares different PPML methods, where HE-HDC-SR (SIMD Representation) denotes the parallel version of HE-HDC.

\ignore{
\figref{lat_breakdown} illustrates the runtime per image of the two approaches in batch inference, where the polynomial degree $N$ (same as the batch size for the batched case) is varied from $2^{11}$ to $2^{13}$. The runtime of HE-HDC is compared with the per image runtime of the batched case, where the total latency is divided by the corresponding batch size. (Refer to \secref{} for details of the experimental setup).
Some things to write about.
}

While the parallel version accelerates HE computation, there is not much difference in the encryption time (of query HVs) and the message size (affecting client-server communication time). As a result, the encryption overhead, which used to be less than 10\% in HE-HDC, now becomes a performance bottleneck, accounting for $\sim$95\% of the client runtime in HE-HDC-SR, which we address next.



\section{Our Proposed Method: \sysname}



\subsection{Client-side HE}

\sysname is a \textit{client-side HE-based PPML method} designed to reduce the overall runtime and network transmission overhead for batch HDC inference.
To reduce the HE encryption overhead, we propose to perform the similarity search operation \emph{homomorphically} on the client as illustrated in \figref{overview}(b).
The similarity search operation requires both class HVs and query HVs, both of which we must protect from the other party's seeing. Therefore, if we move the similarity search operation to the client side, the class HVs must be encrypted while the query HVs need not be any more. This approach allows us to entirely eliminate the encryption time for query HVs. Although class HVs must be encrypted, the encryption time for class HVs is negligible due to their significantly smaller number compared to query HVs.


Our proposed \sysname (HDC with encrypted parameters) can be summarized as follows:
\begin{itemize}
\item Instead of encrypting query HVs into ciphertext, we encrypt class HVs, thereby  eliminating query HV encryption time. 
\item Homomorphic similarity search is performed on the client instead of the server. Moving the similarity search to the client adds to the client runtime but only very slightly. 
\item Reduced communication: instead of a client sending encrypted query HVs, a client sends only the encrypted similarity scores, which is $D$ times smaller in size than query HVs, where $D$ is in the 4K$\sim$10K range. The encrypted class HVs also need to be sent, but only once.
\end{itemize}

The speedup of our proposed approach over the previous HE-HDC mainly comes from three factors: the use of the parallel approach (reducing the HE computation time), the client-side HE (reducing the encryption time and communication time as well as message size), and parameter optimization through our extensive design space exploration.

\subsection{Security Analysis of \sysname}
To keep the model protected from the client, class hypervectors are encrypted into ciphertexts whereas query hypervectors remain as plaintext. 
This client-side approach is valid in the original threat model, since the class hypervectors are not exposed to the users and the user's private data is never sent to the server. We note that the inference result (e.g., object detection result) is revealed to the server, which is not an issue in applications such as face-based authentication assuming that the HE server is also the authentication server, but could be an issue depending on the application scenario.  On the other hand, the revelation of inference result to the server is not a violation of the threat model, and it is hard for the server to recover the original data from the similarity score alone.

It is argued \cite{privhd} that even if query HVs are revealed, raw image can be protected by employing quantization and pruning of query HVs.  Our case is significantly more challenging to break (i.e., recover raw images), since we only reveal the similarity score, not even query HVs. Furthermore, we send a quantized version of similarity scores to the server by virtue of using an integer HE scheme (see \secref{subsection_quant}), which should sufficiently protect the raw images in our case.


\subsection{Applications of \sysname}
\label{section_hdod}

\ignore{\begin{figure}
  \myincludegraphics[width=\linewidth]{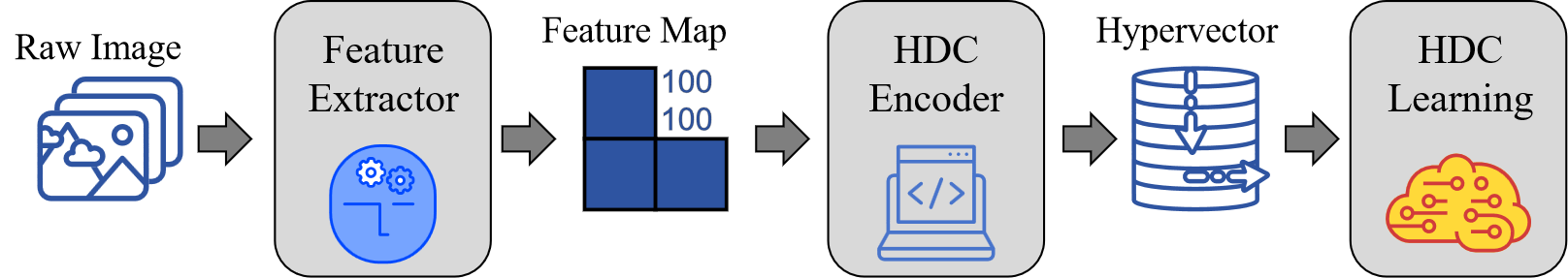}{HDC-based object detection.}
\end{figure}}


\ppheader[]{Encrypted Model Distribution}  This is very similar to a typical PPML use case, where a client has a query, which is sent to the server, and the server performs inference and returns the result to the client.  In our case, the server provides a client with an encrypted model, so that inference would be performed on client's resources.  Also, at the end of inference, the client must communicate with the server to get the plaintext result, which enables the server to limit the usage of the model (e.g., \#inferences per day to curb adversarial attack) just as in conventional PPML.

\ppheader{Privacy-Preserving Authentication}  One difference of the client-side HE from the conventional PPML is that the server can see the plaintext result.  This property can be exploited in applications like face recognition, where the server wants to ensure that the client is a legitimate user by using face image recognition.  Face image may be considered to be private information.  In client-side PPML, the client does not reveal face image but only provides the similarity score, from which the server can learn only the inference result, i.e., if the client is a legitimate user.

\ppheader{Driving Example (HDC-based object detection)}
As a concrete example, we consider the following HDC-based object detection in the rest of the paper.  
While the dataset may vary (e.g., face, emotion), our HDC-based object detection follows these three steps:
\begin{enumerate}
\item Starting with an image frame, we employ a sliding window technique to generate a set of overlapping image patches.
\item For each of the image patches, we utilize a HOG feature extractor to create a corresponding feature map.
\item The generated feature maps are sent to the HDC model, which encodes each feature map to a query hypervector, and calculates the cosine similarity score between a query hypervector and class hypervectors. Based on the similarity score, the model determines whether there is an object in the window as well as the class of the object. 
\end{enumerate}

One limitation of current HDC is the lack of demonstrated state-of-the-art inference accuracy except for some domains, particularly on challenging tasks that require deeper networks.
To alleviate this issue, we employ a feature extractor called HOG, 
which not only helps achieve higher accuracy on HDC but also reduces the computation and memory requirement of an HDC encoder, since extracted data (i.e., HOG feature map) is smaller in size than raw data.

\section{\sysname Design Space Exploration}

\begin{figure}
  \centering
  \includegraphics[width=0.8\linewidth]{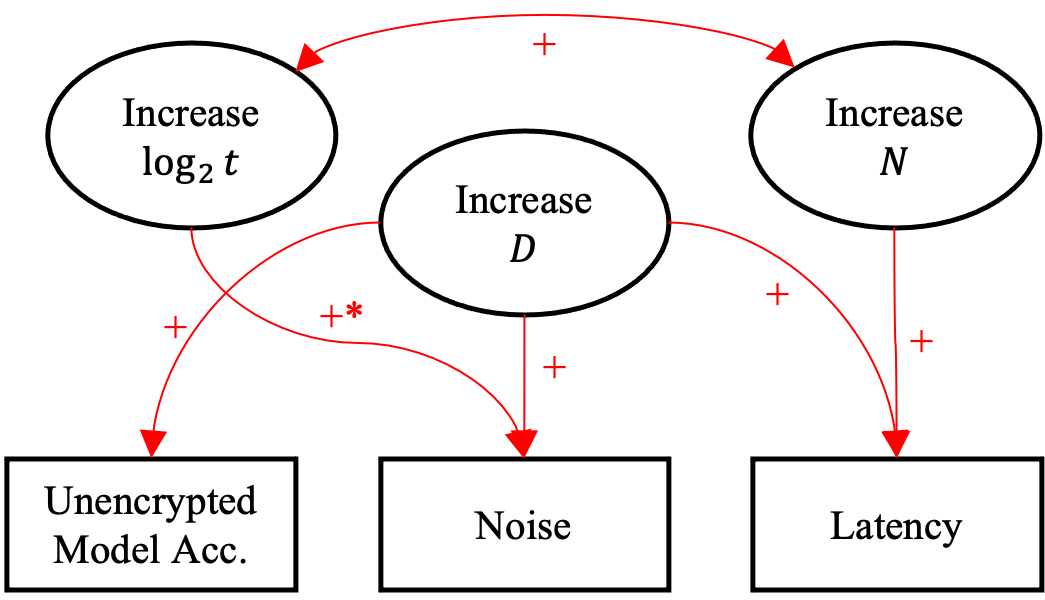} \\
 \parbox{.95\linewidth}{\footnotesize \textit{*Note.~} Increasing $\log_2 t$ does not directly increase noise but reduces the noise \textit{budget}, and hence is similar in effect to increasing the noise level.}
  \caption{Relationship among \sysname parameters.}
  \label{dependency}
\end{figure}

\figref{dependency} illustrates the dependency among the three parameters we explore and their impact.  While HE runtime is mainly determined by $N$ (the polynomial degree of ciphertext), reducing $N$ can also negatively impact the arithmetic precision of HE operations, since $N$ determines the (maximum) $\log_2 t$.
Also, the HDC dimension $D$ heavily influences latency and accuracy. Using larger $D$ helps improve the inference accuracy of unencrypted HDC models; however, it also increases HE noise due to the increased number of homomorphic multiplications and additions during dot-product computation, negatively impacting encrypted inference accuracy. Thus, all three parameters must be explored together in order to find the global optimum.

\subsection{HE-Aware Quantization}
\label{subsection_quant}


\begin{figure}
  \myincludegraphics[width=0.8\linewidth]{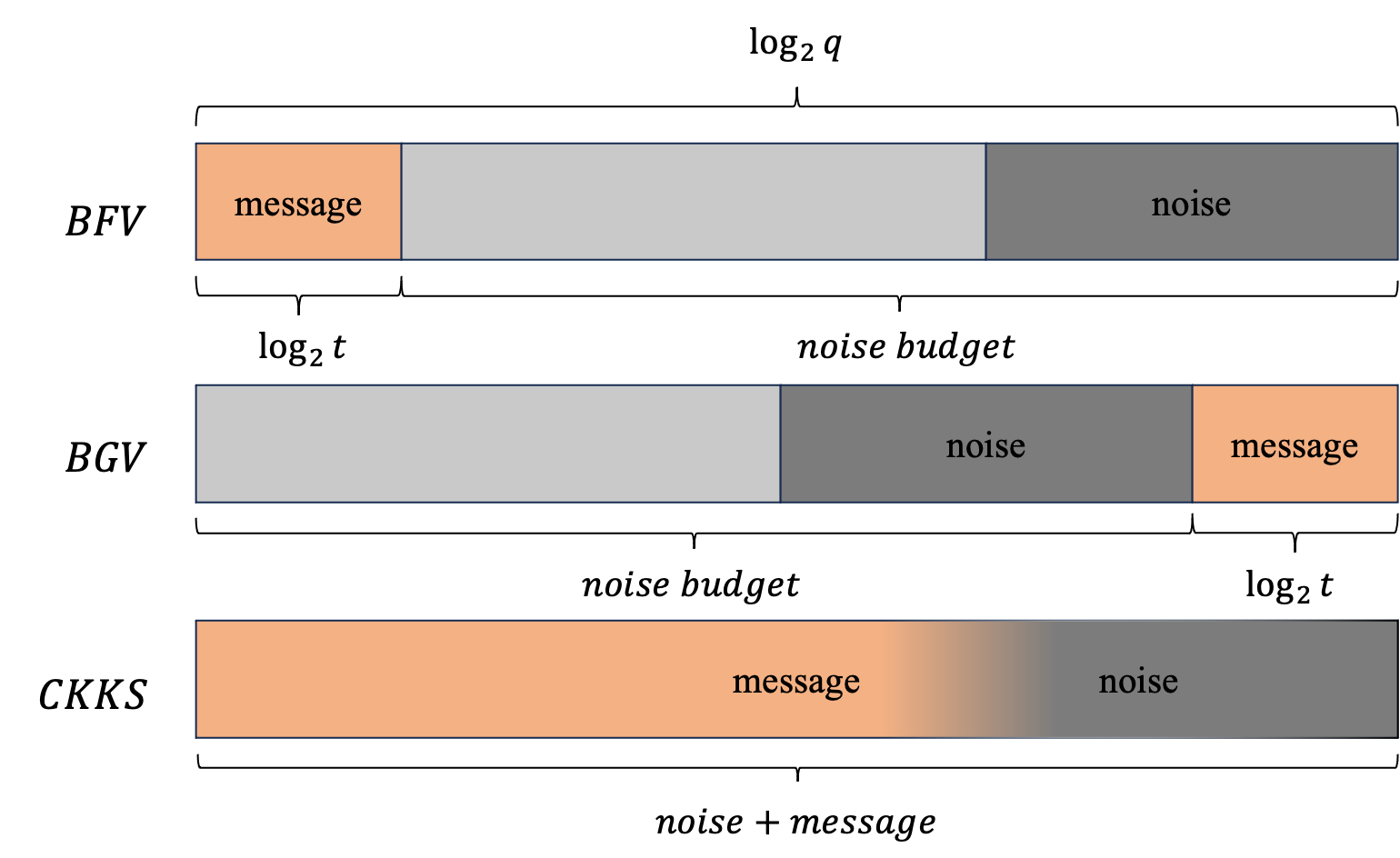}{Encryption method comparison between BFV, BGV, and CKKS.}
  \vspace{-5mm}
\end{figure}

\subsubsection{Quantization Opportunity in BGV/BFV}
We use integer HE schemes (BGV/BFV), which have advantages for our application.  First, they are faster than CKKS when the number of homomorphic multiplications is low \cite{fhebench}. Second, they allow for a more tight control of arithmetic precision than CKKS, which enables further optimization through quantization. 

\figref{scheme} illustrates how message is encrypted into ciphertext in different HE schemes. In CKKS, fixed-point message and noise overlap in the ciphertext where the message can make full use of the $\log_2 q$ bits of precision but the least significant bits may be corrupted due to noise. BGV/BFV scheme reserves for message a certain number of bits (i.e., the plaintext modulo's bit-length $\log_2 t$), leaving the rest of the precision as the noise budget. (Since homomorphic operations gradually increase noise level, the noise budget should be large enough to ensure correct decryption.)
The noise budget can be increased by either lowering the message precision $\log_2 t$ or increasing $N$ (increasing $N$ also increases ciphertext modulo's bit-length $\log_2 q$). 
Since increasing $N$ impacts latency, we seek to reduce message precision $\log_2 t$ to accelerate homomorphic operations without compromising security level.

\subsubsection{BGV/BFV-Aware Quantization}
To maintain the HDC model accuracy at extremely low precision, we apply quantization to both query hypervectors and class hypervectors.
Our quantization approach is similar to quantization-aware training (QAT) in DNNs \cite{lsq,oh21cvpr}, except that whereas DNN quantization, typically, only reduces multiplication precision, our HE-aware quantization requires that \textit{every} computation result be within $\log_2 t$ bits, including those of addition/accumulation. Specifically, to ensure that no overflow happens during addition, we introduce a \emph{scaling operation} similar as introduced in \cite{azat} after every $g$ HE additions, where $g$ is a design parameter (we use 512). 
Note that though ciphertext division is not supported in BGV/BFV, plaintext division is supported if there exists a modular multiplicative inverse of a divisor for a given plaintext modulo $t$, which allows for the use of scaling operations during inference of our quantized model.
The scale parameters are determined through exhaustive search using an encrypted model, considering only powers of two that are divisible for modulo $t$.




\subsection{Exploration Algorithm}
\label{subsection_design_opt}

\ignore{
\jwnote{make this flowchart shorter or make algorithm}
\begin{figure}
  \myincludegraphics[width=0.5\linewidth]{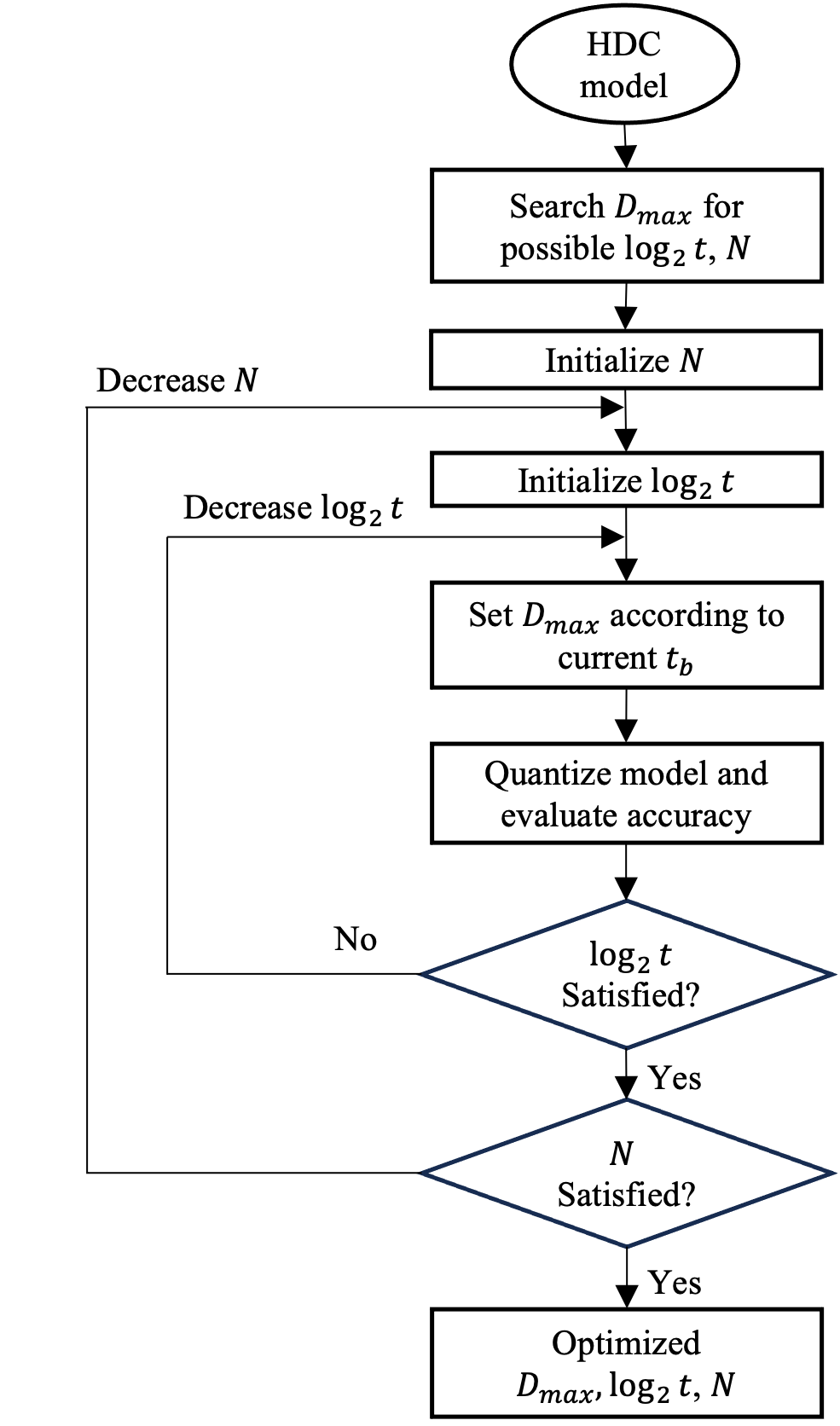}{DSE flow.}
\end{figure}
}

\begin{algorithm}[tb]
\caption{Find the best configuration [$N$, $\log_2 t$, $D$]}
\label{alg}
\begin{algorithmic}[1]
\Procedure{FindBestConfig}{$N_0$}
\State Initialize $N \leftarrow N_0$
\State $T$ $\leftarrow$ getMaxAllowedT($N$) \hfill // $T$ represents $\log_2 t$
\State $D \leftarrow$ \textsc{findDmax}($N$, $T$)
\State $accuracy \leftarrow$ trainHEmodel($N$, $T$, $D$)
\While{$T >$ getMaxAllowedT($N/2$)} 
    \While{$accuracy \geq \textit{acceptable\_acc}$}
        \State Save [$N$, $T$, $D$]
        \State $T \leftarrow T - 1$
        \State $D \leftarrow$ \textsc{findDmax}($N$, $T$)
        \State $accuracy \leftarrow$ trainHEmodel($N$, $T$, $D$)
    \EndWhile
    \State $N \leftarrow N/2$
\EndWhile
\State \Return the last saved [$N$, $T$, $D$] 
\EndProcedure

\Procedure{findDmax}{$N$, $T$}
\For{$i$ \textbf{in} range(1000)}
    \State $D_i \leftarrow 0$  \hfill // initializing $D_{\max}$ to 0
    \While{evalHEcomputation($N$, $T$, $D_i+1$) is correct} 
        \State $D_i \leftarrow D_i+1$   \hfill // updating $D_{\max}$
    \EndWhile
\EndFor
\State \Return  average $\{ D_{i} \}$
\EndProcedure
\end{algorithmic}
\end{algorithm}

A na\"ive approach would be to exhaustively search all parameter combinations, which can be very expensive because (i) the number of combinations can be quite large due to the large number of choices for $D$ and, to a lesser degree, $\log t$, and (ii) estimating accuracy requires model training, which depends on all three parameters. 

Our main ideas are (i) pruning on $N$ and $\log t$, and (ii) eliminating the exploration on $D$.
First, we start from the highest-accuracy parameters (largest $N$ and largest $\log t$), and keep searching for the least parameter values that generate acceptable model accuracy (lines 2--15 of \algref{alg}).
For polynomial degree $N$, only some specific values of $\log t$ are allowed, the maximum of which is obtained by \texttt{getMaxAllowedT}.
Once the model accuracy becomes unacceptable, we need not explore any further with smaller parameters, which allows us to prune the remaining design space.

Second, we choose the largest $D$ allowed cryptographically. 
This is because larger $D$ helps improve the model accuracy while having low impact on latency. In our \sysname, the client latency consists of HE compute time and encoding time with the latter typically being greater, but doubling $D$ doubles the HE compute time only, not the encoding time. Thus, increasing $D$ has limited impact on latency. 
However, increasing $D$ too much can increase the noise level of dot-product computation beyond the noise budget, in which case correct decryption is not guaranteed. Due to the random nature of HE noise, it is hard to obtain the upper bound for $D$ analytically. Thus we find this empirically (\textsc{findDmax}) by evaluating a ciphertext-plaintext vector dot-product for given $D$, $N$, and $\log t$ parameters (line 20), which is repeated many times to obtain the mean value. 

Compared with the na\"ive approach, our exploration algorithm can reduce the number of trainings (line 11) from $O(\mathcal{N} \mathcal{T} \mathcal{D})$ to $O(\mathcal{N} \mathcal{T})$, where $\mathcal{N}, \mathcal{T}, \mathcal{D}$ are the number of choices for $N, \log t, D$ parameters, respectively. The complexity reduction is very significant because $D$ can take any integer in about 1K$\sim$10K.

\section{Experiments}

\subsection{Experimental Setup}
To evaluate \sysname we have implemented the HE part of \sysname using the Pyfhel library \cite{pyfhel}. We use the HDC encoder and the iterative HDC training method of \cite{onlinehd}, extended with a custom quantizer function and HOG. The experiments have been run on a single AMD Ryzen 3970X CPU without multithreading (the same machine configuration is used for both server and client). The previous HDC-based PPML work, HE-HDC \cite{hehdc} is also tested using the same system. We report the \textit{total latency} of each PPML method, which is the sum of the client-side latency and the server-side latency. The total latency includes the encoding of plaintexts, encryption of ciphertexts, and the HE computation latency. In a realistic PPML scenario the network communication latency should also be considered; therefore, we also compare the message size. 


\subsection{HDC Model Accuracy}
\label{subsection_hdc-opt}
We evaluate the unencrypted HDC model accuracy of \sysname on the Face \cite{face} and Emotion \cite{emotion} datasets, both of which are face classification datasets that could be used for object detection applications. 
For a fair comparison with previous PPML works that are crafted to maximize performance on the MNIST dataset, we also present the model accuracy of \sysname and HE-HDC on MNIST. Key statistics of the datasets is listed in \tabref{table_datasets}.

We compare the following cases: 
(i) CryptoNets \cite{cryptonets:icml16} and LoLa \cite{lola}, which share the same DNN model structure, 
(ii) HE-HDC \cite{hehdc}, 
and (iii) EP-HDC. 
EP-HDC uses the same HDC encoder as in HE-HDC, proposed in \cite{onlinehd}, but EP-HDC has an extra HOG feature extractor at the front. \tabref{table_hog} summarizes the accuracy of the methods.
We observe that by employing HOG, \sysname with $D \geq 4096$ shows significantly better accuracy on the Emotion dataset than the CryptoNets/LoLa network.


\begin{table}[t]
  \centering
  \caption{Datasets}
  \label{table_datasets}
  \begin{tabular}{c|ccc}
    \toprule
    & Image size & \#classes     &  \#images \\
    \midrule
    Face   \cite{face} & 1024 $\times$ 1024 & 2 & 40,172 \\
    MNIST  \cite{deng2012mnist} & 28 $\times$ 28 & 10 & 60,000 \\
    Emotion \cite{emotion}& 48 $\times$ 48 & 7 & 36,685 \\
    \bottomrule
  \end{tabular}
\end{table}

\begin{table}[h]
  \centering
  \caption{Unencrypted Model Accuracy Comparison}
   \label{table_hog}
      \resizebox{0.9\linewidth}{!}{  
\begin{tabular}{cc|ccc}
    \toprule
\multirow{2}{*}{Model}         & & \multicolumn{3}{c}{Accuracy (\%)}\\
                               & D  & \multicolumn{1}{c}{Face} & MNIST& \multicolumn{1}{c}{Emotion} \\
                                   \midrule
CryptoNets\cite{cryptonets:icml16} / LoLa \cite{lola}&  & 100.000& \textbf{98.950}& 45.597   \\
            \midrule
\multirow{3}{*}{HE-HDC \cite{hehdc}}        & 1K &           95.375            & 94.110 & 36.528                      \\
                               & 4K & 97.600                    & 96.590 & 39.050                      \\
                               & 8K & 95.825                    & 97.160 & 39.774                      \\
                                       \midrule 
\multirow{3}{*}{\sysname (proposed)} & 1K & 100.000                    & 97.710 & 44.734                      \\
                               & 4K & 100.000                   & 98.390 & 47.910                      \\
                               & 8K & 100.000                    & 98.260 & \textbf{48.899}      \\
                                   \bottomrule
\end{tabular}}
\vspace{-3mm}
\end{table}



\subsection{Effect of Our Client-Side Approach}
To demonstrate that our method does not impose computational burden on the client but rather reduces it, we compare the conventional server-side HE (HE-HDC-SR) vs.\ our \sysname in terms of the \emph{client runtime} on the Face dataset, since the server runtime is obviously reduced by our method.
The results are summarized in \tabref{t:clientrt}. For our method, the client runtime comprises two parts, the HE compute latency and the plaintext encoding (EncPt) latency. In the case of HE-HDC-SR, the client runtime is the ciphertext encryption and decryption latency. 

\begin{table}[]
  \caption{Message size and client runtime in server-side vs.\ client-side HE using Face dataset}
  \label{t:clientrt}
  \centering
\resizebox{0.9\linewidth}{!}{%
\begin{tabular}{c@{~~}c@{~~}c|rc}
\toprule
\multirow{2}{*}{Method}                  & \multirow{2}{*}{N}        & \multirow{2}{*}{D} & Message  & Client Runtime (ms)\\  
  &  &  & Size (B) & (= EncPt + HE Compute)\\ \midrule

\multirow{4}{*}{\shortstack[c]{HE-HDC-SR \\ (Server-side HE)}} & \multirow{2}{*}{$2^{11}$} & 1K                 &    31.86M  &                          ~351.95 \\
                             &                           & 2K                 &            63.71M     &                           ~702.86\\
                                         & \multirow{2}{*}{$2^{12}$} & 1K                 &        131.2M  &                           1163.95 \\
                                   &                           & 2K                 &             262.4M   &                      2303.22 \\
                        \midrule
\multirow{4}{*}{\shortstack[c]{\sysname \\ (Client-side HE)}}          & \multirow{2}{*}{$2^{11}$} & 1K                 &          31.1K&                         ~43.99 (= 27.76 + 16.23) \\
                                         &                           & 2K                 &   31.1K   &                             ~87.72 (= 55.47 + 32.25)\\
                                         & \multirow{2}{*}{$2^{12}$} & 1K                 &      121.1K    &                            105.27 (= 54.97 + 50.30)\\
                                         &                           & 2K                 &        121.1K &                           204.60 (= 103.7 + 100.9) \\
                                         \bottomrule
\end{tabular}}
\end{table}

The results show that HE encryption time (of HE-HDC-SR) dominates the client runtime of \sysname (including both plaintext encoding and HE compute) by many times.  Using the client-side approach is on average $9.1\times$ faster than server-side approach. The total runtime, including the server runtime, is improved by $8.1\times$ for $N=2^{11}$ and $11.2\times$ for $N=2^{12}$. Along with the latency improvement, the HE message size that has to be transmitted between the client and the server is also reduced in the client-side approach by $D$ times, which is very significant.

\subsection{Effect of HE Noise}
\begin{figure}
    \myincludegraphics[width=\linewidth]{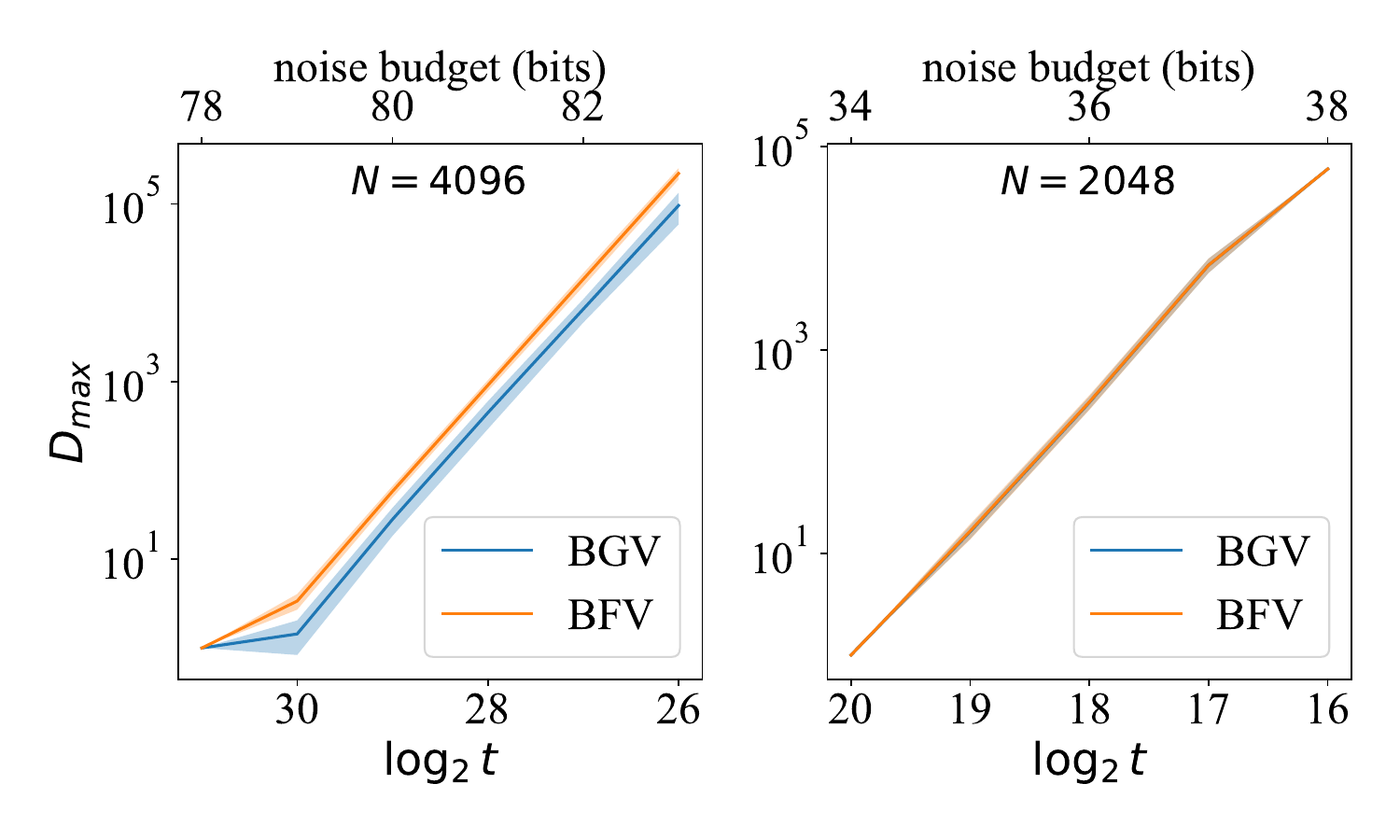}{$D_{\max}$ for various $N$ and $\log_2 t$. The width of shaded region represents standard deviation.}
\end{figure}

\label{sec:noise_result}
As performing more homomorphic operations accumulates more HE noise, we observe that the hypervector dimension $D$ is the most crucial factor in determining the noise level of our system. 
Due to the random nature of HE noise, there is no deterministic upper-bound for hypervector dimension. Instead, we use the \textsc{findDmax} procedure in \algref{alg}, the result of which is presented in \figref{noise_growth}. 
The graph suggests a clear trade-off between $\log_2 t$ and $D_{\max}$, in which decreasing $\log_2 t$ increases $D_{\max}$ exponentially. We also observe that BFV has better noise management than BGV, allowing for more HE operations. Based on this observation, we use BFV in the following experiments.
\ignore{
\begin{table}[]
  \centering
    \caption{Latency and Throughput Results for different parameters}
\begin{tabular}{cc|rrr|r}
    \toprule
\multirow{2}{*}{N}    & \multirow{2}{*}{D} & \multicolumn{3}{c}{Latency}   & \multirow{2}{*}{Throughput} \\
                      &                    & HDC Encoding & HE Inference & Total  &                             \\
                          \midrule
\multirow{4}{*}{$2^{11}$} & 1K                 & 99.89     &  45.68  &145.57 &7034.38                   \\
                      & 2K                 &137.37     &  90.72  &228.09 &4489.38                \\
                      & 4K                 & 205.04    &  179.81 & 384.85&2660.79         \\
                      & 8K                 &335.96   & 366.84 & 702.80&1457.04                     \\
                      \midrule
\multirow{4}{*}{$2^{12}$} & 1K                 & 197.93   & 104.92&302.85 & 6762.42               \\
                      & 2K                 &257.96   & 210.81 &468.77 &4368.88                    \\
                      & 4K                 & 385.58& 416.23 &801.81&2554.22                \\
                      & 8K                 &624.98& 833.86 & 1458.84&1403.86   \\
                          \bottomrule
\end{tabular}
\end{table}
}
\subsection{Effect of Optimizations in \sysname}
The initial implementation of \sysname uses 16-bit quantization for the class and query hypervectors, where a polynomial degree of at least $N=2^{13}$ is required. Following the optimization method from \secref{subsection_design_opt}, we have optimized \sysname for the Face and Emotion datasets, the result of which is summarized in \tabref{opti_results}. For the Face dataset, quantizing query hypervectors to 6 bits and class hypervectors to 2 bits can reduce the polynomial degree to $N=2^{11}$. In the case of Emotion dataset, model accuracy of 44.97\% can be reached with $N=2^{11}$. This optimization improves the HE latency by $6.65\times$ for the Face dataset without any accuracy loss and $9.74\times$ for the Emotion dataset with minimal accuracy drop. 

\begin{table}[]
\centering
\caption{Effect of Optimizations in \sysname}
\label{opti_results}
\begin{tabular}{c|cc|cc}
\toprule
                & \multicolumn{2}{c|}{Face \cite{face}} & \multicolumn{2}{c}{Emotion \cite{emotion}} \\
                & Initial  & Optimized  & Initial   & Optimized   \\
                \midrule
$N$               & 8192         & 2048       & 8192          & 2048        \\
$\log_2 t$      & 60           & 17         & 60            & 16          \\
$D$               & 1024         & 1024       & 8192          & 5632        \\

$Q, C$*    & 16, 16       & 6, 2       & 16, 16        & 8, 8         \\
\midrule
Accuracy (\%)   & 100.00          & 100.00       & 48.899        & 44.970         \\
HE Latency (ms) & 292.69       & \textbf{43.99}      & 2306.47       & \textbf{236.85}     \\
 \bottomrule
 \multicolumn{5}{l}{*$Q$-bit quantization for query HVs and $C$-bit quantization for class HVs.}\\
\end{tabular}
\vspace{-5mm}
\end{table}

\begin{table}
  \caption{Comparison with Previous HDC-based PPML Methods}
    \centering
  \label{table_previous_work} 
\begin{tabular}{c|cc|cc}
\toprule
Dataset          &\multicolumn{2}{c|}{Face} & \multicolumn{2}{c}{Emotion}  \\
Method   & HE-HDC    & \sysname   &  HE-HDC  & \sysname \\ 
\midrule
Batch Size       &     1     &  1024      &    1     & 292    \\
Accuracy (\%)    & 97.23  &   \textbf{100.0}    & 39.05       & \textbf{44.97}     \\
Message Size &    96.2\,KB  &   \textbf{31.1\,KB}  & 96.2\,KB   & \textbf{31.1\,KB}    \\
Latency (s)      &    \textbf{0.086}  &   0.144   & \textbf{0.106}  &0.317    \\
Throughput       & 11.57  &\textbf{7117.04}     &  9.43  & \textbf{912.13}        \\

\bottomrule
\end{tabular}
\end{table}

\subsection{Comparison with Previous HE-based PPML}

\begin{table}
  \caption{Comparison with Previous PPML Methods (MNIST)}
    \centering
  \label{table_previous_work_dnn} 
\begin{tabular}{c|ccc}
\toprule
Method     & CryptoNets & LoLa & \sysname \\ 
\midrule
Batch Size            &    4096      &  1   & 204    \\
Accuracy (\%)     & \textbf{98.95}        &   \textbf{98.95} &  98.39     \\
Message Size  &    367.5MB & 11.72MB  & \textbf{31.1\,KB}    \\
Latency (s)    &    250*    & 2.2*$^\dagger$   &\textbf{0.341}    \\
Throughput      & 16.38*    & 0.45*$^\dagger$    & \textbf{598.24}        \\

\bottomrule
\multicolumn{4}{l}{*Quoted from the original papers. ~$^\dagger$With multithreading enabled.}
\end{tabular}
\vspace{-5mm}
\end{table}

\tabref{table_previous_work} provides a comparison with HE-HDC \cite{hehdc}, a HDC-based PPML method, on the Face and Emotion datasets, while \tabref{table_previous_work_dnn} is a comparison with DNN-based PPML methods, CryptoNets \cite{cryptonets:icml16} and LoLa \cite{lola}, on the MNIST dataset. We have implemented HE-HDC, modifying it for different datasets (see \tabref{table_datasets}).  
Latency is the total latency, and throughput (= batch size / latency) is the number of images (or the number of sliding windows for Face and Emotion datasets) that can be processed per second. 
The results indicate that compared to HE-HDC, \sysname has superior accuracy and much higher throughput (96.7$\sim$612\X) but slightly longer latency, which is due to the HOG feature extractor as well as the different message representation. 
Compared to DNN-based methods, \sysname has both higher throughput (36.52$\sim$1068\X) and faster latency (6.45$\sim$733\X), with marginally lower accuracy.
These results suggest that our \sysname can be a viable solution for PPML applications where both privacy and high performance are crucial.

\section{Related Work}
\subsubsection*{Privacy-Preserving Machine Learning}
Dowlin \etal \cite{cryptonets:icml16} present \emph{CryptoNets}, neural networks that can be applied to encrypted data to make accurate predictions while maintaining data privacy and security.
\emph{Faster CryptoNets} \cite{chou2018faster} accelerates the homomorphic evaluation by employing a pruning and quantization approach that leverages sparse representations.
Low-Latency CryptoNets \cite{lola} changes data representation throughout computation in a novel way to reduce latency while maintaining accuracy and security. 
Hesamifard \etal \cite{ehsan2019} replace nonlinear activation functions with low-degree polynomials.
Recently, an HDC-based approach \emph{HE-HDC} \cite{hehdc} was proposed, which can be implemented using HE-friendly operations only, thus improving the latency of PPML considerably. 

\subsubsection*{Hyperdimensional Computing}
NeuralHD \cite{neuralhd} is the first HDC algorithm with a dynamic and regenerative encoder for adaptive learning, and can enhance learning capability and robustness by identifying insignificant dimensions and regenerating those dimensions. 
VoiceHD \cite{voicehd} is an efficient and hardware-friendly speech recognition technique using HD computing, which maps preprocessed voice signals in the frequency domain to hypervectors and combines them to compute class hypervectors. 
AdaptHD \cite{adapthd} is an adaptive retraining method for HD computing, which introduces the idea of learning rate to HD computing and proposes a hybrid approach to update the learning rate considering both iteration and data dependency.
Instead of simple hypervector averaging, OnlineHD \cite{onlinehd} updates a model differently depending on the model prediction result, enabling iterative training and potentially boosting performance of HDC models.
While HOG has been applied to HDC in prior work \cite{facehd}, our design is different in that we apply HOG directly to raw images rather than to hypervectors.

Lastly, various works on accelerating HE have been proposed, ranging from ASIC-based methods \cite{sharp,ark}, to processing-in-memory based methods such as \cite{nttpim}. 

\section{Conclusion}
We presented \sysname, a novel HDC-based PPML that can deliver much higher performance than any previous HE-based PPML methods. Our novel features include (i) client-side HE which reduces the encryption overhead by 9.1\X (ii) BFV/BGV-aware quantization and (iii) a novel parameter optimization method that improves HE latency by 6.65$\sim$9.74\X. Our results demonstrate that our \sysname can outperform state-of-the-art DNN-based PPML methods with 36.52$\sim$1068\X higher throughput and 6.45$\sim$733\X faster latency at $<$1\% accuracy degradation.



%% file: main.bbl
\begin{thebibliography}{10}
\providecommand{\url}[1]{#1}
\csname url@samestyle\endcsname
\providecommand{\newblock}{\relax}
\providecommand{\bibinfo}[2]{#2}
\providecommand{\BIBentrySTDinterwordspacing}{\spaceskip=0pt\relax}
\providecommand{\BIBentryALTinterwordstretchfactor}{4}
\providecommand{\BIBentryALTinterwordspacing}{\spaceskip=\fontdimen2\font plus
\BIBentryALTinterwordstretchfactor\fontdimen3\font minus \fontdimen4\font\relax}
\providecommand{\BIBforeignlanguage}[2]{{%
\expandafter\ifx\csname l@#1\endcsname\relax
\typeout{** WARNING: IEEEtran.bst: No hyphenation pattern has been}%
\typeout{** loaded for the language `#1'. Using the pattern for}%
\typeout{** the default language instead.}%
\else
\language=\csname l@#1\endcsname
\fi
#2}}
\providecommand{\BIBdecl}{\relax}
\BIBdecl

\bibitem{cryptonets:icml16}
R.~Gilad-Bachrach, N.~Dowlin, K.~Laine, K.~Lauter, M.~Naehrig, and J.~Wernsing, ``{CryptoNets}: Applying neural networks to encrypted data with high throughput and accuracy,'' in \emph{ICML}.\hskip 1em plus 0.5em minus 0.4em\relax PMLR, 2016, pp. 201--210.

\bibitem{lola}
A.~Brutzkus, R.~Gilad-Bachrach, and O.~Elisha, ``Low latency privacy preserving inference,'' in \emph{ICML}.\hskip 1em plus 0.5em minus 0.4em\relax PMLR, 2019, pp. 812--821.

\bibitem{hehdc}
J.~Park, C.~Quan, H.~Moon, and J.~Lee, ``Hyperdimensional computing as a rescue for efficient privacy-preserving machine learning-as-a-service,'' in \emph{ICCAD}, Oct. 2023.

\bibitem{face}
P.~Kottarathil, ``Face mask lite dataset,'' \url{https://www.kaggle.com/prasoonkottarathil/face-mask-lite-dataset}, 2020.

\bibitem{emotion}
``Emotion detection dataset,'' \url{https://www.kaggle.com/datasets/ananthu017/emotion-detection-fer}.

\bibitem{kanerva2009hyperdimensional}
P.~Kanerva, ``Hyperdimensional computing: An introduction to computing in distributed representation with high-dimensional random vectors,'' \emph{Cognitive computation}, vol.~1, pp. 139--159, 2009.

\bibitem{rahimi2016hyperdimensional}
A.~Rahimi, S.~Benatti, P.~Kanerva, L.~Benini, and J.~M. Rabaey, ``Hyperdimensional biosignal processing: A case study for emg-based hand gesture recognition,'' in \emph{2016 IEEE International Conference on Rebooting Computing (ICRC)}.\hskip 1em plus 0.5em minus 0.4em\relax IEEE, 2016, pp. 1--8.

\bibitem{7838428}
H.~Li, T.~F. Wu, A.~Rahimi, K.-S. Li, M.~Rusch, C.-H. Lin, J.-L. Hsu, M.~M. Sabry, S.~B. Eryilmaz, J.~Sohn, W.-C. Chiu, M.-C. Chen, T.-T. Wu, J.-M. Shieh, W.-K. Yeh, J.~M. Rabaey, S.~Mitra, and H.-S.~P. Wong, ``Hyperdimensional computing with {3D} {VRRAM} in-memory kernels: Device-architecture co-design for energy-efficient, error-resilient language recognition,'' in \emph{IEEE IEDM}, 2016.

\bibitem{thomas2021theoretical}
A.~Thomas, S.~Dasgupta, and T.~Rosing, ``A theoretical perspective on hyperdimensional computing,'' \emph{Journal of Artificial Intelligence Research}, vol.~72, pp. 215--249, 2021.

\bibitem{neuralhd}
Z.~Zou, Y.~Kim, F.~Imani, H.~Alimohamadi, R.~Cammarota, and M.~Imani, ``Scalable edge-based hyperdimensional learning system with brain-like neural adaptation,'' in \emph{Proceedings of the International Conference for High Performance Computing, Networking, Storage and Analysis}, 2021.

\bibitem{voicehd}
M.~Imani, D.~Kong, A.~Rahimi, and T.~Rosing, ``{VoiceHD}: Hyperdimensional computing for efficient speech recognition,'' in \emph{2017 IEEE International Conference on Rebooting Computing (ICRC)}, 2017, pp. 1--8.

\bibitem{adapthd}
M.~Imani, J.~Morris, S.~Bosch, H.~Shu, G.~D. Micheli, and T.~Rosing, ``{AdaptHD}: Adaptive efficient training for brain-inspired hyperdimensional computing,'' in \emph{2019 IEEE Biomedical Circuits and Systems Conference (BioCAS)}, 2019, pp. 1--4.

\bibitem{onlinehd}
A.~Hernández-Cano, N.~Matsumoto, E.~Ping, and M.~Imani, ``{OnlineHD}: Robust, efficient, and single-pass online learning using hyperdimensional system,'' in \emph{DATE}, 2021, pp. 56--61.

\bibitem{ckks}
J.~H. Cheon, A.~Kim, M.~Kim, and Y.~Song, ``Homomorphic encryption for arithmetic of approximate numbers,'' in \emph{Advances in Cryptology--ASIACRYPT 2017}.\hskip 1em plus 0.5em minus 0.4em\relax Springer, 2017, pp. 409--437.

\bibitem{bgv}
Z.~Brakerski, C.~Gentry, and V.~Vaikuntanathan, ``Fully homomorphic encryption without bootstrapping,'' Cryptology ePrint Archive, Paper 2011/277, 2011.

\bibitem{bfv}
J.~Fan and F.~Vercauteren, ``Somewhat practical fully homomorphic encryption,'' Cryptology ePrint Archive, Paper 2012/144, 2012.

\bibitem{reagen2021cheetah}
B.~Reagen, W.-S. Choi, Y.~Ko, V.~T. Lee, H.-H.~S. Lee, G.-Y. Wei, and D.~Brooks, ``Cheetah: Optimizing and accelerating homomorphic encryption for private inference,'' in \emph{HPCA}.\hskip 1em plus 0.5em minus 0.4em\relax IEEE, 2021, pp. 26--39.

\bibitem{gazelle}
C.~Juvekar, V.~Vaikuntanathan, and A.~Chandrakasan, ``Gazelle: A low latency framework for secure neural network inference,'' in \emph{Proceedings of the 27th USENIX Conference on Security Symposium}, ser. SEC'18, 2018, p. 1651–1668.

\bibitem{privhd}
B.~Khaleghi, M.~Imani, and T.~Rosing, ``Prive-hd: Privacy-preserved hyperdimensional computing,'' in \emph{Proceedings of the 57th ACM/EDAC/IEEE Design Automation Conference}, ser. DAC '20.\hskip 1em plus 0.5em minus 0.4em\relax IEEE Press, 2020.

\bibitem{fhebench}
L.~Jiang and L.~Ju, ``Fhebench: Benchmarking fully homomorphic encryption schemes,'' \emph{arXiv preprint arXiv:2203.00728}, 2022.

\bibitem{lsq}
S.~K. Esser, J.~L. McKinstry, D.~Bablani, R.~Appuswamy, and D.~S. Modha, ``Learned step size quantization,'' in \emph{International Conference on Learning Representations}, 2019.

\bibitem{oh21cvpr}
S.~Oh, H.~Sim, S.~Lee, and J.~Lee, ``Automated log-scale quantization for low-cost deep neural networks,'' in \emph{Proceedings of the IEEE/CVF Conference on Computer Vision and Pattern Recognition (CVPR)}, June 2021, pp. 742--751.

\bibitem{azat}
A.~Azamat, J.~Park, and J.~Lee, ``Squeezing accumulators in binary neural networks for extremely resource-constrained applications,'' in \emph{ICCAD}.\hskip 1em plus 0.5em minus 0.4em\relax ACM, 2022.

\bibitem{pyfhel}
A.~Ibarrondo and A.~Viand, ``Pyfhel: Python for homomorphic encryption libraries,'' in \emph{Proceedings of the 9th on Workshop on Encrypted Computing \& Applied Homomorphic Cryptography}, 2021, pp. 11--16.

\bibitem{deng2012mnist}
L.~Deng, ``The mnist database of handwritten digit images for machine learning research,'' \emph{IEEE Signal Processing Magazine}, vol.~29, no.~6, pp. 141--142, 2012.

\bibitem{chou2018faster}
E.~Chou, J.~Beal, D.~Levy, S.~Yeung, A.~Haque, and L.~Fei-Fei, ``{Faster CryptoNets}: Leveraging sparsity for real-world encrypted inference,'' \emph{CoRR}, vol. abs/1811.09953, 2018.

\bibitem{ehsan2019}
E.~Hesamifard, H.~Takabi, and M.~Ghasemi, ``Deep neural networks classification over encrypted data,'' in \emph{CODASPY '19}.\hskip 1em plus 0.5em minus 0.4em\relax ACM, 2019.

\bibitem{facehd}
\BIBentryALTinterwordspacing
M.~Imani, A.~Zakeri, H.~Chen, T.~Kim, P.~Poduval, H.~Lee, Y.~Kim, E.~Sadredini, and F.~Imani, ``Neural computation for robust and holographic face detection,'' in \emph{Proceedings of the 59th ACM/IEEE Design Automation Conference}, ser. DAC '22.\hskip 1em plus 0.5em minus 0.4em\relax New York, NY, USA: Association for Computing Machinery, 2022, p. 31–36. [Online]. Available: \url{https://doi.org/10.1145/3489517.3530653}
\BIBentrySTDinterwordspacing

\bibitem{sharp}
\BIBentryALTinterwordspacing
J.~Kim, S.~Kim, J.~Choi, J.~Park, D.~Kim, and J.~H. Ahn, ``Sharp: A short-word hierarchical accelerator for robust and practical fully homomorphic encryption,'' in \emph{Proceedings of the 50th Annual International Symposium on Computer Architecture}, ser. ISCA '23.\hskip 1em plus 0.5em minus 0.4em\relax New York, NY, USA: Association for Computing Machinery, 2023. [Online]. Available: \url{https://doi.org/10.1145/3579371.3589053}
\BIBentrySTDinterwordspacing

\bibitem{ark}
\BIBentryALTinterwordspacing
J.~Kim, G.~Lee, S.~Kim, G.~Sohn, M.~Rhu, J.~Kim, and J.~H. Ahn, ``Ark: Fully homomorphic encryption accelerator with runtime data generation and inter-operation key reuse,'' in \emph{Proceedings of the 55th Annual IEEE/ACM International Symposium on Microarchitecture}, ser. MICRO '22.\hskip 1em plus 0.5em minus 0.4em\relax IEEE Press, 2023, p. 1237–1254. [Online]. Available: \url{https://doi.org/10.1109/MICRO56248.2022.00086}
\BIBentrySTDinterwordspacing

\bibitem{nttpim}
J.~Park, S.~Lee, and J.~Lee, ``Ntt-pim: Row-centric architecture and mapping for efficient number-theoretic transform on pim,'' in \emph{2023 60th ACM/IEEE Design Automation Conference (DAC)}, 2023, pp. 1--6.

\end{thebibliography}
